\documentstyle[12pt]{article}
\begin{document}
\setlength{\baselineskip}{24pt}
\begin{center}{
{\large \bf PHASE SPACE REPRESENTATION \\
FOR OPEN QUANTUM SYSTEMS \\
WITHIN THE LINDBLAD THEORY}\\
\vskip 1truecm
A. Isar$^{\dagger\ddagger}$, A. Sandulescu$^\dagger$, and W. Scheid$^
\ddagger$\\
$^\dagger${\it Department of Theoretical Physics, Institute of
Atomic Physics \\
Bucharest-Magurele, Romania }\\
$^\ddagger${\it Institut f\"ur Theoretische Physik der
Justus-Liebig-Universit\"at \\
Giessen, Germany }\\
}
\end{center}
\vskip 1truecm
\begin{abstract}
\vskip 0.5truecm

The Lindblad master equation for an open quantum system with a Hamiltonian
containing an arbitrary potential is written 
as an equation for the Wigner distribution function
in the phase space representation. The time
derivative of this function is given by a sum of three parts: the classical
one, the quantum corrections and the contribution due to the opening of the
system. In the particular case of a harmonic oscillator, quantum
corrections do not exist.

\end{abstract}

\section{Introduction}

In the last two decades, more and more interest has arisen about the problem of
dissipation in quantum mechanics and the search for a consistent description
of open quantum systems [1-4]. Because dissipative processes imply
irreversibility and, therefore, a preferred direction in time, it is generally
thought that quantum dynamical semigroups are the basic tools to introduce
dissipation in quantum mechanics. A general form of the generators of
such semigroups was given by Lindblad [5-7]. This formalism has been studied
for the case of damped harmonic oscillators [8,9] and applied to various
physical phenomena, for instance, to the damping of collective modes in deep
inelastic collisions in nuclear physics [10-12] and to the interaction of a
two-level atom with a single mode of the electromagnetic field
[13] (for a recent review see Ref. [14]).

The quasiprobability distribution functions have proven to be of great use in
the study of quantum mechanical systems. They are useful not only as
calculational tools, but can also provide insights into the connection between
classical and quantum mechanics.
The first of these quasiprobability distributions was introduced by Wigner
[15] to study quantum corrections to classical statistical mechanics. The
Wigner distribution function has found many applications primarily in 
statistical
mechanics or in purely quantum mechanical problems, but also in areas such as
quantum chemistry and quantum optics, collisions [16], quantum
chaos [17,18]
quantum fluid dynamics and for some aspects of density functional
theory [19].
Quantum optics has given rise to a number of quasiprobability distributions,
the most well-known being the Glauber $P$ distribution [20,21].  More
recently extensive use has been made of the generalized $P$ distributions
[22,23].

In the previous papers [24,25] the applicability of quasiprobability
distributions to the Lindblad theory was explored. In order to have a simple
formalism we studied the master equation of the one-dimensional damped harmonic
oscillator as example of an open quantum system. This equation was transformed
into Fokker-Planck type equations for $c$-number quasiprobability distributions
associated with the density operator and a comparative study of the Glauber,
antinormal ordering and Wigner representations was made. We then solved
these Fokker-Planck equations, subject to different types of initial
conditions.

The increased activity in heavy ion physics has led to a greater interest in
phase space distributions. Currently models as the intranuclear cascade
model, "hot spot" model and hydrodynamical models, which
emphasize the collective and
transport behaviour, are especially suitable for an examination in this
framework [16]. We could also mention the nuclear reaction theory formulated
in the language of the Wigner distribution function [26], the
use of a modified
Vlasov equation for the description of intermediate energy or high energy heavy
ion collisions [27-29], the quantum molecular dynamics approach to
investigate the fragment formation and the nuclear equation of state in heavy 
ion
collisions [30] and the transport phenomena in dissipative heavy ion
collisions [31].

The aim of the present paper is to use the quantum mechanical phase space
Wigner distribution for a
one-dimensional particle in a general potential in the framework of the
Lindblad theory
for open quantum systems. Since the Wigner distribution is a quantum
generalization of the Boltzmann one-particle distribution, we can obtain 
an analogue of the Boltzmann-Vlasov equation for open quantum systems.

The content of this paper is arranged as follows. In Sec. 2 we review very
briefly the Lindblad model for the theory of open quantum systems, namely we
formulate the basic equation of motion for the density operator.
In Sec. 3 we
discuss the quantum mechanical phase space Wigner distribution and its
properties, which are quite useful in applications. In Sec. 4  we derive a 
Boltzmann-Vlasov type equation from the Lindblad master equation
for the time
dependence of the Wigner distribution. Finally, we discuss and summarize our
results in Sec. 5.

\section{Quantum mechanical Markovian master equation}

A possibility for an irreversible behaviour in a finite system is to avoid
the unitary time development and to consider non-Hamiltonian systems.
If $S$ is a limited set of macroscopic degrees of freedom and 
$R$ the
set of non-explicitly described degrees of freedom of a large system $S+R,$ the
simplest dynamics for $S$ which describes an irreversible process is a
semigroup of transformations introducing a preferred direction in time.

In Lindblad's axiomatic formalism of introducing dissipation in quantum
mechanics, the usual von Neumann-Liouville equation ruling the time evolution
of closed quantum systems is replaced by the following Markovian master
equation for the density operator $\hat\rho(t)$ in the Schr\"odinger picture
[5]:
$${d\Phi_{t}(\hat\rho)\over dt}=L(\Phi_{t}(\hat\rho)). \eqno (2.1)$$
Here, $\Phi_{t}$ denotes the dynamical semigroup describing the irreversible
time evolution of the open system in the Schr\"odinger representation and $L$
is the infinitesimal generator of the dynamical semigroup $\Phi_t$. Using the
Lindblad theorem which gives the most general form of a bounded, completely
dissipative generator $L$, we obtain the explicit form of the most general
quantum mechanical master equation of Markovian type:
$${d\hat\rho(t)\over dt}=-{i\over\hbar}[\hat H,\hat\rho(t)]+{1\over 2\hbar}
\sum_{j}([\hat V_{j}\hat\rho(t),\hat V_{j}^\dagger ]+[\hat
V_{j},\hat\rho(t)\hat V_{j}^\dagger ]).\eqno (2.2)$$
Here $\hat H$ is the Hamiltonian operator of the system and $\hat V_{j},$
$\hat V_{j}^\dagger $ are bounded operators on the Hilbert space $\cal H$ 
of the Hamiltonian.
We should like to mention that the Markovian master equations found in the
literature are of this form after some rearrangement of terms, even for
unbounded generators. We make the basic assumption that the general
form (2.2) of the master equation with a bounded generator is also valid for
an unbounded generator. Since we study the one-dimensional problem, we
consider that the operators $\hat H,\hat V_{j},\hat V_{j}^\dagger $
are functions of the basic observables $\hat p$ and $\hat q$ of the
one-dimensional quantum mechanical system (with $[\hat q,\hat p]=i\hbar\hat I,
$ where $\hat I$ is the identity operator on $\cal H)$.
For simplicity we consider that the operators $\hat V_{j},$ $\hat
V_j^\dagger $ are first degree polynomials in $\hat p$ and $\hat q$.
Since in the linear space of the first degree polynomials in $\hat p$ and
$\hat q$ the operators $\hat p$ and $\hat q$ give a basis, there exist only two
${\bf C}$-linear independent operators $\hat V_{1},\hat V_{2}$ which can be
written in the form of
$$\hat V_{j}=a_{j}\hat p+b_{j}\hat q,~j=1,2,\eqno (2.3)$$
with $a_{j},b_{j}$ complex numbers [6]. The constant term is omitted
because its contribution to the generator $L$ is equivalent to terms in
$\hat H$
linear in $\hat p$ and $\hat q$ which for simplicity are chosen to be zero, 
so that $\hat H$ is
taken of the form
$$\hat H=\hat H_0+{\mu\over 2}(\hat p \hat q+\hat q \hat p), ~~
\hat H_0={1\over 2m}\hat p^2+U(\hat q),
\eqno (2.4)$$
where $U(\hat q)$ is the potential energy and $m$ is the mass of the particle.
With these choices and with the notations
$$D_{qq}={\hbar\over 2}\sum_{j=1,2}{\vert a_{j}\vert}^2,
  D_{pp}={\hbar\over 2}\sum_{j=1,2}{\vert b_{j}\vert}^2,
D_{pq}=D_{qp}=-{\hbar\over 2}{\rm Re}\sum_{j=1,2}a_{j}^*b_{j},
\lambda=-{\rm Im}\sum_{j=1,2}a_{j}^*b_{j},\eqno(2.5)$$
where $a_j^*$ and $b_j^*$ denote the complex conjugate of $a_j,$
and $b_j^*$ respectively,
the master equation (2.2) takes the following form [8,14]:
$${d\hat\rho \over dt}=-{i\over \hbar}[\hat H_0,\hat\rho]+{i(\lambda-\mu)
\over 2\hbar}
[\hat p,\hat\rho\hat q+\hat q\hat\rho]-{i(\lambda+\mu)\over 2\hbar
}[\hat q,\hat\rho \hat p+\hat p\hat\rho]$$
$$-{D_{pp}\over {\hbar}^2}[\hat q,[\hat q,\hat\rho]]-{D_{qq}\over {\hbar}^2}
[\hat p,[\hat p,\hat\rho]]+
{2D_{pq}\over {\hbar}^2}[\hat p,[\hat q,\hat\rho]]. \eqno (2.6)$$
Here the quantum mechanical diffusion coefficients $D_{pp},D_{qq},$ $D_{pq}$
and the friction constant $\lambda$ satisfy the following fundamental
constraints [8,14]:
$${\rm i})~D_{pp}>0,~{\rm ii})~D_{qq}>0,
~{\rm iii})~D_{pp}D_{qq}-{D_{pq}}^2\ge {\lambda}^2{\hbar}^2/4. \eqno (2.7)$$

The necessary and sufficient condition for $L$ to be translationally
invariant is $\mu=\lambda$ [6,8,14]. In the following general values
for $\lambda$ and $\mu$ will be considered.

\section{Wigner distribution function}

In the phase space picture, the quantum corrections become transparent
and a smooth transition from quantum to classical physics is
encountered. This picture is particularly suitable for obtaining quantum
mechanical results in situations where a good initial approximation
comes from the classical result and also for deriving classical limits
of quantal processes.

A quantum mechanical particle is described by a density matrix
$\hat\rho$ and the average of a function of the position and momentum
operators $\hat A(\hat p,\hat q)$ (we consider the one-dimensional case) is
$$<\hat A>={\rm Tr}(\hat A\hat\rho),\eqno(3.1)$$
or, in terms of a quasiprobability distribution $\phi(p,q),$
$$<\hat A>=\int dp \int dq A(p,q)\phi(p,q), \eqno (3.2) $$
where the function $A(p,q)$ can be derived from the operator $\hat A(\hat p,
\hat q)$ by a well-defined correspondence rule [32].
This allows one to cast quantum mechanical results into a form in which
they resemble classical ones. In the case where $\phi$ in Eq. (3.2) is
chosen to be the Wigner distribution, then the correspondence between
$A(p,q)$ and $\hat A$ is that proposed by Weyl [33], as was first
demonstrated by Moyal [34].

The Wigner distribution function, which represents the Weyl transform
of the density operator, is defined through the partial Fourier
transform of the off-diagonal elements of the density matrix:
$$W(p,q)={1\over\pi\hbar}\int dy<q-y\vert\hat\rho\vert q+y> e^{2ipy/\hbar}
\eqno(3.3)$$
and satisfies the following properties:

${\rm ~~i})~W(p,q)$ is real, but cannot be everywhere positive,

${\rm ~ii}) \int dp W(p,q)=<q |\hat\rho|q>,$~~~~~~~~~~~~~~~~~~~~~~~~~~~~~~~~
~~~~~~~~~~~~~~~~~~~~~~~~~~~~~~~~~~~~(3.4)

$~~~~\int dq W(p,q)=<p|\hat\rho|p>,$~~~~~~~~~~~~~~~~~~~~~~~~~~~~~~~~~~~~~~~
~~~~~~~~~~~~~~~~~~~~~~~~~~~~~~(3.5)

$~~~~\int dp\int dq W(p,q)={\rm Tr}\hat\rho=1,$~~~~~~~~~~~~~~~~~~~~~~~~~~~~~
~~~~~~~~~~~~~~~~~~~~~~~~~~~~~~~~~~~(3.6)

${\rm iii})~W(p,q)$ is Galilei invariant and invariant with respect to space
and time reflec-

~~~~~tions,

${\rm iv})$ In the force free case the equation of motion is the classical
(Liouville) one
$${\partial W\over\partial t}=-{p\over m}{\partial W\over\partial q}.
\eqno(3.7)$$
Wigner has shown that any real distribution function as long as it
satisfies properties ${\rm ii})$ assumes also negative values for some $p$
and $q.$

The classical phase space function $A(p,q)$ corresponding to the
quantum operator $\hat A$ is given by the so-called Wigner-Moyal transform:
$$A(p,q)=\int dz e^{ipz/\hbar}<q-{1\over 2}z\vert\hat A\vert q+{1\over 2}z>,
\eqno(3.8)$$
so that  $\int\int dp dq A(p,q)=2\pi\hbar{\rm Tr}\hat A.$
Clearly, Eq. (3.3) is a special case of Eq. (3.8) for the density
operator, i.e. $\hat A=\hat\rho$ and $W(p,q)$ is the phase space
function which corresponds to the operator $\hat\rho/2\pi\hbar.$ The
function $A(p,q)$ is known as the Wigner equivalent of the operator $\hat A.$
The relation which expresses the Wigner equivalent $F$ of the product
of operators $\hat F=\hat A\hat B$ in terms of the Wigner equivalents of
the individual operators $\hat A$ and $\hat B$ is the following:
$$F(p,q)=A(p,q)(\exp{\hbar\Lambda\over
2i})(p,q)=B(p,q)(\exp(-{\hbar\Lambda\over 2i}))A(p,q),\eqno(3.9)$$
where $\Lambda$ (essentially the Poisson bracket operator) is given by
$$\Lambda=\overleftarrow{\partial\over\partial p}\overrightarrow{\partial\over
\partial q}-\overleftarrow{\partial\over\partial q}\overrightarrow{\partial
\over\partial p}\eqno(3.10)$$
and the arrows indicate in which direction the derivatives act.
From Eq. (3.9) the Wigner equivalent of the commutator $[\hat A,\hat B]$ 
follows directly:
$$([\hat A,\hat B])(p,q)=-2iA(p,q)(\sin{\hbar\Lambda\over 2})B(p,q).
\eqno(3.11)$$
Next we recall the time dependence of the Wigner distribution function
$W(p,q,t)$ for a closed system [32]. 
Instead of
the von Neumann-Liouville master equation
we have the following quantum Liouville equation which determines the time
evolution of the Wigner phase space distribution function (Wigner equivalent of
the density operator):
$$i\hbar {\partial W\over\partial t}=H(p,q)(\exp{\hbar\Lambda\over
2i})W(p,q)-W(p,q)(\exp{\hbar\Lambda\over 2i})H(p,q),\eqno(3.12)$$
or, cf. Eq. (3.11),
$$\hbar {\partial W\over\partial t}=-2H(p,q)(\sin{\hbar\Lambda\over
2})W(p,q),\eqno(3.13)$$
where $H(p,q)$ is the Wigner equivalent of the Hamiltonian operator $\hat H$
of the system. 
Note that if we take the
$\hbar\to 0$ limit of this equation, we obtain the classical Liouville equation
(Vlasov equation).

\section{The equation of motion for the Wigner distribution of open systems}

In two previous papers [8,24] the evolution equations for the Wigner
function were obtained for the damped harmonic oscillator within the
Lindblad theory for open quantum systems. In this Section we obtain
the equation for the Wigner function of a
one-dimensional system with an arbitrary potential. In other words, we
obtain, by means of the Wigner function, the phase space representation
of the general Lindblad master equation.

The time evolution
of the Wigner distribution function corresponding to the
Lindblad master equation
(2.2), can be obtained from Eq. (3.12) by adding in the right-hand side
the Wigner equivalent corresponding to the opening part of Eq. (2.2),
i.e. the sum of commutators. By using formulas (3.9) and (3.11), we obtain
the following evolution equation for the Wigner distribution:
$${\partial W\over\partial t}=-{2\over\hbar}H(\sin{\hbar\Lambda\over2})W$$
$$+{1\over 2\hbar}\sum_j\left [2V_j(\exp{\hbar\Lambda\over 2i})W(\exp{\hbar
\Lambda\over 2i})V_j^*-V_j^*(\exp{\hbar\Lambda\over 2i})V_j(\exp{\hbar
\Lambda\over 2i})W-W(\exp{\hbar\Lambda\over 2i})V_j^*(\exp{\hbar
\Lambda\over 2i})V_j\right ],
\eqno(4.1)$$
where $V_j$ and $V_j^*$
are the Wigner equivalents of the operators $\hat V_j$ and
$\hat V_j^\dagger,$ respectively, and $V_j^*$ is the complex
conjugate of $V_j.$ If the operators $\hat
V_j$ are taken of the form (2.3), then Eq. (4.1) becomes the evolution
equation for the Wigner function corresponding to the master equation (2.6):
$$ {\partial W\over \partial t}=-{2 \over\hbar}H(\sin{\hbar\Lambda\over 2})W+
\lambda{\partial\over\partial q}(qW)+\lambda{\partial\over\partial p}(pW)$$
$$+ D_{qq}{\partial^2 W\over \partial q^2}+D_{pp}{\partial^2 W
\over \partial p^2}+2D_{pq}{\partial^2 W \over \partial p \partial q}.
\eqno (4.2) $$
Here it is easily to remark that the first term on the right-hand
side generates the
evolution in phase space of a closed system and gives the
Poisson bracket and the higher derivatives containing the
quantum contribution, and the following terms represent the
contribution from the opening (interaction with the environment).

For a Hamiltonian operator of the form (2.4) and by assuming a
Taylor expansion of the potential $U,$ i.e. $U(q)$ is an analytic
function, this equation takes the form:
$${\partial W\over\partial t}=-{p\over m}{\partial W\over\partial q}
+{\partial U\over\partial q}{\partial W\over\partial p}+\sum_{n=1}
^\infty(-1)^n{(\hbar)^{2n}\over2^{2n}(2n+1)!}{\partial^{2n+1}U(q)\over\partial
q^{2n+1}}{\partial^{2n+1}W\over\partial p^{2n+1}}+{\cal L}W,\eqno(4.3)$$
where we have introduced the notation 
$${\cal L}\equiv (\lambda-\mu){\partial\over\partial q}(qW)+(\lambda+\mu)
{\partial\over\partial p}(pW)
+ D_{qq}{\partial^2 W\over \partial q^2}+D_{pp}{\partial^2 W
\over \partial p^2}+2D_{pq}{\partial^2 W \over \partial p \partial q}.
\eqno (4.4) $$
If Eq. (4.3) had only the first two terms on the right-hand side, $W$
would evolve along the classical flow in phase-space. The terms
containing $\lambda$ and $\mu$ are the dissipative terms. They
modify the flow of $W$ and cause a
contraction of each volume element in phase space. The terms containing 
$D_{pp}, D_{qq}$ and $D_{pq}$ are the diffusive terms and produce an
expansion of the volume elements.
The diffusion terms are responsible for noise and also for the
destruction of interference, by erasing the structure of the
Wigner function on small scales. The sum term (the power series),
together with the first two terms make up the unitary part of
the evolution. Hence, up to corrections of order $\hbar^2,$
unitary evolution corresponds to approximately classical
evolution of the Wigner function. It is partly for this reason
that the evolution of a quantum system is most conveniently
undertaken in the Wigner representation. The higher corrections
can often be assumed as negligible and give structures on small
scales. There are, however, important examples where they
cannot be neglected, e.g., in chaotic systems (see the
discussion below).
From Eq. (4.3) it is clear that, as a
consequence of the quantum correction terms with higher
derivatives, the Wigner function of a non-linear
system does not follow the classical Liouville flow. The higher
derivative terms are generated by the nonlinearities in the potential $U(q).$
   
The question of the classical limit of the Wigner equation is of more than
formal interest [35]. There are two well-known limits in which
Eq. (4.3) can go over into a classical equation:
1) when $U$ is at most quadratic in $q$ and 
2) when $\hbar\to 0.$

Because of the extra diffusion terms, we get yet a third
classical limit: In the limit of large $D_{pp},$ the diffusive
smoothing becomes so effective that it damps out all the
momentum-derivatives in the infinite sum and Eq. (4.3)
approaches the Liouville equation with diffusion, an equation of
Fokker-Planck type. This is an example of how
macroscopic objects start to behave classically (decoherence), since the
diffusion coefficients are roughly proportional to the size of
these objects. Thus an object will evolve according to classical
dynamics if it has a strong interaction with its environment.
The connection between decoherence and transition from quantum
to classical in the framework of the Lindblad theory for open
quantum systems will form the subject of another paper [36].

In the following we shall consider Eq. (4.3) for some particular cases.

1) In the case of a free particle, i.e. $U(q)=0,$
Eq. (4.3) takes the form:
$${\partial W\over\partial t}=-{p\over m}{\partial W\over\partial q}
+{\cal L}W. \eqno(4.5)$$


2) In the case of a linear potential $U=\gamma q$ (where
for example $\gamma=mg$ for the free fall or $\gamma=eE$ for the
motion in a uniform electric field), one gets
$${\partial W\over\partial t}=-{p\over m}{\partial W\over\partial q}
+\gamma{\partial W\over\partial p}+{\cal L}W. \eqno(4.6)$$


3) In the case of the harmonic oscillator with 
$U=m\omega^2 q^2/2,$ Eq. (4.3) takes the form:
$${\partial W\over \partial t}=-{p \over m}{\partial W \over
\partial q}+m \omega^2q{\partial W\over \partial p}+{\cal L}W. \eqno(4.7)$$
An analogous equation can be obtained in the case of the motion
on an inverted parabolic potential $U(q)=-m\kappa^2 q^2/2.$ 
Since the drift coefficients are linear in the variables $p$ and $q$ and
the diffusion coefficients are constant with respect to $p$ and $q,$ 
Eqs. (4.5)-(4.7)
describe an Ornstein-Uhlenbeck process [37-39].
Eqs. (4.5)-(4.7) are exactly equations
of the Fokker-Planck type. Eq. (4.7) was extensively studied is
the literature [8,14,25] and   
it represents an
exactly solvable model. It should be stressed that not every function $W(p,q,0)
$ on the phase-space is the  Wigner transform of a density operator. Hence, the
quantum mechanics appears now in the restrictions imposed 
on the initial condition $W(p,q,0)$ for Eq. (4.3). The most
frequently used choice for $W(p,q,0)$ is a Gaussian function and Eqs.
(4.5)-(4.7) preserve this Gaussian type,
i.e., $W(p,q,t)$ is always a Gaussian function in time, so that the 
differences between
quantum and classical mechanics are completely lost in this
representation of the master equation.

4) In the case of an exponential potential $U(q)=\alpha\exp(-\beta q),$
the Wigner equation is an infinite order partial differential equation

$${\partial W\over\partial t}=-{p\over m}{\partial W\over\partial q}
-\alpha\beta\exp(-\beta q){\partial W\over\partial p}+\sum_{n=1}
^\infty(-1)^{3n+1}{(\hbar)^{2n}\over2^{2n}(2n+1)!}\alpha\beta^{2n+1}\exp(-\beta
q){\partial^{2n+1}W\over\partial p^{2n+1}}+{\cal L}W,\eqno(4.8)$$
but in the case of a potential of the finite polynomial form
$U(q)=\sum_{n=1}^N a_n q^n,$ the sum keeps only a certain number
of derivate terms.
As an illustration of this remark, we consider an anharmonic oscillator 
with the potential $U_{anh}(q)
=m\omega^2q^2/2
+Cq^4.$ In this case the Wigner equation (4.3) becomes

$${\partial W\over \partial t}=-{p \over m}{\partial W \over
\partial q}+(m \omega^2q+4Cq^3){\partial W\over \partial p}
-C\hbar^2q{\partial^3 W\over\partial p^3}+{\cal L}W. \eqno(4.9)$$
The above equation has one term with third derivative, associated
with the nonlinear potential $U_{anh}.$ In fact, the first three
terms on the right-hand side of Eq. (4.9) give the usual Wigner
equation of an isolated anharmonic oscillator. The third
derivative term is of order $\hbar^2$ and is the
quantum correction. In the classical
limit, when this term is neglected, the Wigner equation becomes
one of the Fokker-Planck type. 

5) For a periodic potential $U(q)=U_0\cos(kq),$ 

$${\partial^{2n+1}U\over\partial
q^{2n+1}}=(-1)^n k^{2n}{\partial U\over\partial
q} \eqno(4.10)$$
and we obtain

$${\partial W\over \partial t}=-{p \over m}{\partial W \over
\partial q}+{\partial U\over\partial q}{\delta W\over \delta p}
+{\cal L}W, \eqno(4.11) $$
where
$${\delta W\over \delta p}=
{W(q,p+\hbar k/2,t)-W(q,p-\hbar k/2,t)\over \hbar k}. \eqno(4.12)$$
Eq. (4.11) takes a simpler form when $\hbar k$ is large compared
to the momentum spread $\Delta p$ of the particle being
considered, i.e. when the spatial extension of the wave packet
representing the particle is large compared to the spatial
period of the potential. Imposing the condition $\hbar k\gg
\Delta p$ on Eq. (4.12), one sees that $\delta W/ \delta p$ is
small for any $p$ that yields an appreciable value of the
Wigner distribution function, i.e. $\delta W/ \delta
p\approx 0$ for all practical purposes [40]. Eq. (4.11) is then reduced
to the equation (4.5) for a free particle moving in an environment.

From the examples 1)--3) we see that for Hamiltonians at most quadratic in 
$q$ and $p,$ the equation of
motion of the Wigner function contains only the classical part
and the contributions from the opening of the system
and obeys classical Fokker-Planck equations of motion
(4.5)-(4.7).  
In general, of course, the potential $U$ has terms of order
higher than $q^2$ and one has to deal with a partial
differential equation of order higher than two or generally of
infinite order.
When the potential deviates only slightly from the harmonic
potential, one can still take the classical limit $\hbar\to 0$
in Eq. (4.3) as the lowest-order approximation to the quantum
motion and construct higher-order approximations that contain
quantum corrections to the classical trajectories using the
standard perturbation technique [40]. 

Eq. (4.3) can also be
rewritten in the form
$${\partial W\over \partial t}=-{p \over m}{\partial W \over
\partial q}+{\partial U_{eff}\over\partial q}{\partial W\over \partial p}
+{\cal L}W, \eqno(4.13) $$
where the effective potential is defined as [40]

$${\partial U_{eff}\over\partial q}{\partial W\over\partial p}=
{\partial U\over\partial q}{\partial W\over\partial p}+\sum_{n=1}
^\infty(-1)^n{(\hbar)^{2n}\over2^{2n}(2n+1)!}{\partial^{2n+1}U(q)\over\partial
q^{2n+1}}{\partial^{2n+1}W\over\partial p^{2n+1}}. \eqno(4.14)$$
Then the phase-space points move under the influence of the effective potential
$U_{eff}.$ We note that Eq. (4.14) indicates that only when at
least an approximate solution for $W$ is known, the effective
potential can be determined. If $U$ does not deviate much from
the harmonic potential, the zeroth-order approximation for $W$
can be taken as that resulting from classical propagation. The
main limitation of the effective potential method is that it can
be applied only to systems whose behaviour is not much different
from that of the harmonic oscillator or the free wave packet,
for example a low-energy Morse oscillator and an almost free
wave packet slightly perturbed by a potential step or barrier
[40]. The effective
potential method may possibly be applied to a collision
system [40]. The quantum dynamics of atomic and
molecular collision processes has been described rather
successfully in the past by means of classical trajectories
[16,40-42], i.e. the collision system is one for which an
approximate solution of $W$ can be considered as known and thus the
perturbative scheme of the effective potential method can be employed.
In Refs. [43-45] similar equations were written for the 
Wigner function for a class of quantum Brownian motion models
consisting of a particle moving in a general potential and
coupled to an environment of harmonic oscillators.

Phase space provides a natural framework to study the
consequences of the chaotic dynamics and its interplay with
decoherence [46-50]. Eq. (4.3) could be applied in order to
investigate general implications of the process of decoherence
for quantum chaos. Since decoherence induces a transition from
quantum to classical mechanics, it can be used to find the connection
between the classical and quantum chaotic systems. In this case
$U(q)$ is the potential of a classically chaotic system, coupled
to the external environment. For this purpose, a particular case
of Eq. (4.3) was utilized by Zurek and Paz [51] in the
special case of the high temperature limit of the environment: 
$\mu=\lambda=\gamma,$ where $\gamma$ is the relaxation rate,
$D_{qq}=0,D_{pq}=0$ and the diffusion coefficient
$D_{pp}\equiv D=2m\gamma k_B T $ ($T$ is the temperature of the
environment). In this model the symmetry between $q$ and $p$ is
broken, and coupling with the environment through position gives
momentum diffusion only [52]. Zurek also
used a similar equation in which the diffusion term is
symmetric, namely $D(\partial_{pp}^2+\partial_{qq}^2)W$ [51].

It is sometimes argued that the power series involving third and
higher derivative terms may be neglected. Paz and Zurek [51] have
argued that the diffusive terms may smooth out the Wigner
function, suppressing contributions from the higher-order terms. 
When these terms can be neglected, the Wigner function evolution
equation (4.3) then becomes
$${\partial W\over \partial t}=-{p \over m}{\partial W \over
\partial q}+{\partial U\over\partial q}{\partial W\over \partial p}
+{\cal L}W. \eqno(4.15)$$
In the previous considered particular case of a thermal bath and if 
$\lambda=\mu=\gamma,$
$D_{qq}=D_{pq}=0,$ $D_{pp}\equiv D=2m\gamma k_B T,$ this equation becomes
a Kramers equation [45,53]:

$${\partial W\over \partial t}=-{p \over m}{\partial W \over
\partial q}+{\partial U\over\partial q}{\partial W\over \partial p}
 + 2\gamma
{\partial\over\partial p}(pW)
+2m\gamma k_B T{\partial^2 W\over\partial p^2}.\eqno(4.16)$$
It posseses the stationary solution [45]
$$W(p,q)=\tilde N\exp\left[-{p^2\over 2mk_B T}-{U(q)\over k_B T}\right], 
\eqno(4.17)$$
where $\tilde N$ is a normalization factor. As was discussed by
Anastopoulos and Halliwell [45], this will be an
admissable solution, i.e., it is a Wigner function only if the
potential is such that $\exp[-U(q)/D]$ is normalizable. This
requires $U(q)\to\infty$ as $q\to\pm\infty$ faster than
$\ln|q|.$ In this case the stationary distribution is the
Wigner transform of a thermal state $\rho=Z^{-1}\exp(-H_0/kT)$
with $Z={\rm Tr}\exp(-H_0/kT)$ for large temperatures. The
general results of [53] then show that all solutions of Eq.
(4.16) approach the stationary solutions (4.17) as $t\to\infty.$
Hence, to the extent that Eq. (4.16) is valid, all initial states
tend towards the thermal state in the long-time limit [45].

\section{Summary and outlook}

The Lindblad theory provides a selfconsistent treatment of damping as a
possible extension of quantum mechanics to open systems. In the present paper
we have shown how the Wigner distribution can be used to solve
the problem of dissipation for some simple systems. From the master
equation we have derived the corresponding Vlasov or Fokker-Planck equations
in the Wigner $W$ representation. The Fokker-Planck equations which we
obtained describe an Ornstein-Uhlenbeck process.
The Wigner representation discussed provides a classical
conceptual framework for
studying quantum phenomena and enables one to employ 
various methods of approximation or expansion used in classical cases
to problems of the quantum domain. By means of the Wigner function
one can calculate mean values of physical quantities that may be of
interest, using a classical-like phase space integration, where position
and momentum are treated as ordinary variables rather than as operators.

The phase-space formulation of quantum mechanics represents an
alternative to the standard wave mechanics formulation of the
Lindblad equation. The main difficulty with the phase-space
formulation is that the time development of the phase-space
distribution (Wigner) function is given in terms of an
infinite-order partial differential equation (see Eq. (4.3)). What
matters, however, is not whether the equation can be solved
exactly, but whether a reasonable set of approximations can be
introduced that make it possible to obtain an approximate but
still accurate solution to the equation. The Wigner phase-space
formulation is particularly useful when the classical
description of the system under consideration is adequate and
quantum corrections to the classical description are only desired for
higher accuracy. 

{\it Acknowledgment.} One of us (A.I.) whould like to express his
sincere gratitude to Professor W. Scheid for the hospitality at the
Institut f\"ur Theoretische Physik in Giessen.

{\large {\bf References}}
\begin{enumerate}
\item
R. W. Hasse, J. Math. Phys. {\bf 16} (1975) 2005

\item
E. B. Davies, {\it Quantum Theory of Open Systems} (Academic Press, New York, 
1976)

\item
H. Dekker, Phys. Rep. {\bf 80} (1981) 1

\item
K. H. Li, Phys. Rep. {\bf 134} (1986) 1

\item
G. Lindblad, Commun. Math. Phys. {\bf 48} (1976) 119

\item
G. Lindblad, Rep. Math. Phys. {\bf 10} (1976) 393

\item
G. Lindblad, {\it Non-Equilibrium Entropy and Irreversibility} (Reidel, 
Dordrecht, 1983)

\item
A. Sandulescu and H. Scutaru, Ann. Phys. (N.Y.) {\bf 173} (1987) 277

\item
A. Sandulescu, H. Scutaru and W. Scheid, J. Phys. A - Math. Gen. {\bf 20}
(1987) 2121

\item
A. Pop, A. Sandulescu, H. Scutaru and W. Greiner, Z. Phys. A {\bf 329} (1988)
357

\item
A. Isar, A. Sandulescu and W. Scheid, Int. J. Mod. Phys. A {\bf 5}
(1990) 1773

\item
A. Isar, A. Sandulescu and W. Scheid, J. Phys. G - Nucl. Part. Phys. {\bf 17}
(1991) 385

\item
A. Sandulescu and E. Stefanescu, Physica A {\bf 161} (1989) 525

\item
A. Isar, A. Sandulescu, H. Scutaru, E. Stefaanescu and W. Scheid, 
Int. J. Mod. Phys. E {\bf 3}
(1994) 635

\item
E. P. Wigner, Phys. Rev. {\bf 40} (1932) 749

\item
P. Carruthers and F. Zachariasen, Rev. Mod. Phys. {\bf 55} (1983) 245

\item
K. Takahasi and N. Saito, Phys. Rev. Lett. {\bf 55} (1985) 645

\item
K. Takahasi, J. Phys. Soc. Jpn. {\bf 55} (1986) 762

\item
{\it The Physics of Phase Space,} ed. by Y. S. Kim and W. W. Zachary (Springer,
Berlin, 1986)

\item
E. J. Glauber, Phys. Rev. Lett. {\bf 10} (1963) 84

\item
E. C. G. Sudarshan, Phys. Rev. Lett. {\bf 10} (1963) 277

\item
P. D. Drummond and C. W. Gardiner, J. Phys. A - Math. Gen. {\bf 13} (1980) 2353

\item
P. D. Drummond, C. W. Gardiner and D. F. Walls, Phys. Rev. A {\bf 24} (1981)
914

\item
A. Isar, W. Scheid and A. Sandulescu, J. Math. Phys. {\bf 32} (1991) 2128

\item
A. Isar, Helv. Phys. Acta {\bf 67} (1994) 436

\item
E. A. Remler, Ann. Phys. (N.Y.) {\bf 136} (1981) 293

\item
H. St\" ocker and W. Greiner, Phys. Rep. {\bf 137} (1986) 278

\item
G. F. Bertsch and S. Das Gupta, Phys. Rep. {\bf 160} (1988) 191

\item
W. Cassing, V. Metag, U. Mosel and K. Niita, Phys. Rep. {\bf 188} (1990) 363

\item
J. Aichelin, Phys. Rep. {\bf 202} (1991) 233

\item
H. Feldmeier, Rep. Prog. Phys. {\bf 50} (1987) 915

\item
M. Hillery, R.F. O 'Connell, M.O. Scully and E.P. Wigner, Phys.Rep.
{\bf 106} (1984) 121

\item
H. Weyl, Z. Phys. {\bf 46} (1927) 1

\item
J. E. Moyal, Proc. Cambridge Phil. Soc. {\bf 45} (1949) 99

\item
E. Heller, J. Chem. Phys. {\bf 65} (1976) 1289

\item
A. Isar and W. Scheid, in preparation 

\item
G. E. Uhlenbeck and L. S. Ornstein, Phys. Rev. {\bf 36} (1930) 823

\item
H. Haken, Rev. Mod. Phys. {\bf 47} (1975) 67

\item
C. W. Gardiner, {\it Handbook of Stochastic Methods} (Springer, Berlin, 1982)

\item
H. W. Lee, Phys. Rep. {\bf 259} (1995) 147

\item
D. L. Bunker, Method. Comput. Phys. {\bf 10} (1971) 287

\item
L. M. Raff and D. L. Thomson, in {\it Theory of Chemical Reaction Dynamics,} 
ed. by M. Baer (CRC, Boca Ratou, 1985) vol. III, p. 1

\item
B. L. Hu, J. P. Paz and Y. Zhang, Phys. Rev. D {\bf 47} (1993) 1576

\item
A. Anderson and J. J. Halliwell, Phys. Rev. D {\bf 48} (1993) 2753

\item
C. Anastapoulos and J. J. Halliwell, Phys. Rev. D {\bf 51}
(1995) 6870

\item
W. H. Zurek, Phys. Rev. D {\bf 24} (1981) 1516; {\it ibid.} {\bf
26} (1982) 1862; E. Joos and H. D. Zeh, Zeit. Phys. B {\bf 59}
(1985) 229

\item
W. H. Zurek, Phys. Today {\bf 44} (Oct. 1991) 36; {\it ibid.} {\bf
46} (Apr. 1993) 81; 
Prog. Theor. Phys. {\bf 89} (1993) 281

\item
J. P. Paz, S. Habib and W. H. Zurek, Phys. Rev. D {\bf 47}
(1993) 488

\item
W. H. Zurek, S. Habib and J. P. Paz, Phys. Rev. Lett. {\bf 70} (1993) 1187

\item
W. H. Zurek, in {\it Frontiers of Nonequilibrium Statistical Mechanics,}
ed. by G. T. Moore and M. O. Scully (Plenum, New York, 1986)

\item
W. H. Zurek and J. P. Paz, Phys. Rev. Lett. {\bf 72} (1994) 2508

\item
A. O. Caldeira and A. J. Leggett, Physica A {\bf 121} (1983) 587;
W. G. Unruh and W. H. Zurek, Phys. Rev. D {\bf 40} (1989) 1071; 
B. L. Hu, J. P. Paz and Y Zhang, Phys. Rev. D {\bf 45} (1992) 2843;
{\it ibid.} {\bf47} (1993) 1576

\item
H. Risken, {\it The Fokker-Planck Equations: Methods of Solution
and Applications,}
(Springer-Verlag, Berlin, 1989)

\end{enumerate}
\end{document}